\documentclass[twocolumn]{aastex631}

\newcommand\lsim{\mathrel{\rlap{\lower4pt\hbox{\hskip1pt$\sim$}}
\raise1pt\hbox{$<$}}}
\usepackage{xcolor}

\shorttitle{IMBH Formation in Galactic Nuclei}
\shortauthors{Rose et al.}

\graphicspath{{./}{figures/}}

\begin{document}

\title{
The Formation of Intermediate Mass Black Holes in Galactic Nuclei}

\correspondingauthor{Sanaea C. Rose}
\email{srose@astro.ucla.edu}

\author{Sanaea C. Rose}
\affiliation{Department of Physics and Astronomy, University of California, Los Angeles, CA 90095, USA}
\affiliation{Mani L. Bhaumik Institute for Theoretical Physics,
University of California, Los Angeles,
CA 90095, USA}

\author{Smadar Naoz}
\affiliation{Department of Physics and Astronomy, University of California, Los Angeles, CA 90095, USA}
\affiliation{Mani L. Bhaumik Institute for Theoretical Physics,
University of California, Los Angeles,
CA 90095, USA}

\author{Re'em Sari}
\affiliation{Racah Institute for Physics, The Hebrew University, Jerusalem 91904, Israel}

\author{Itai Linial}
\affiliation{Racah Institute for Physics, The Hebrew University, Jerusalem 91904, Israel}

\begin{abstract}
Most stellar evolution models predict that black holes (BHs) should not exist above approximately $50-70$~M$_\odot$, the lower limit of the pair-instability mass gap. However, recent LIGO/Virgo detections indicate the existence of BHs with masses at and above this threshold. We suggest that massive BHs, including intermediate mass black holes (IMBHs), can form in galactic nuclei through collisions between stellar-mass black holes and the surrounding main-sequence stars. Considering dynamical processes such as collisions, mass segregation, and relaxation, we find that this channel can be quite efficient, forming IMBHs as massive as $10^4$~M$_\odot$. This upper limit assumes that (1) the BHs accrete a substantial fraction of the stellar mass captured during each collision and (2) that the rate at which new stars are introduced into the region near the SMBH is high enough to offset depletion by stellar disruptions and star-star collisions. We discuss deviations from these key assumptions in the text. Our results suggest that BHs in the pair-instability mass gap and IMBHs may be ubiquitous in galactic centers. This formation channel has implications for observations. Collisions between stars and BHs can produce electromagnetic signatures, for example, from x-ray binaries and tidal disruption events. Additionally, formed through this channel, both black holes in the mass gap and IMBHs can merge with the supermassive black hole at the center of a galactic nucleus through gravitational waves. These gravitational wave events are extreme and intermediate mass ratio inspirals (EMRIs and IMRIs, respectively).
\end{abstract}


\section{Introduction} \label{sec:intro}

The recently detected gravitational wave source GW190521 \citep{GW190521a+20,GW190521b+20} produced an intermediate mass black hole of approximately $142$~M$_\odot$. This event may have also had a $85$~M$_\odot$ progenitor, which falls within the pair-instability mass gap that limits stellar black holes (BHs) to no more than $\lsim 50$~M$_\odot$ \citep[e.g.,][]{Heger+03,Woosley+17}\footnote{Note that the exact lower and upper limits may be sensitive to metallicity of the progenitor \citep[e.g.,][]{Woosley+17,Spera+17,Limongi+18,Sakstein+20,Belczynski+20,Renzo+20,Vink+21}.}. Similarly, the merger products of GW150914, GW170104, and GW170814 fall within the mass gap \citep[e.g.,][]{GW150914,GW170104,GW170814}. BH mergers that form second generation BHs and, in some cases, intermediate mass BHs (IMBHs), these gravitational wave (GW) events can occur in globular clusters, young stellar clusters, or the field \citep[e.g.,][]{Rodriguez+18,Rodriguez+19,Fishbach+20,Mapelli+21,Mapelli+21Cosmic,DiCarlo+19,DiCarlo+21,DallAmico+21,ArcaSedda+21}. However, IMBHs are not limited to these locations and may reside in galactic nuclei as well. Several studies propose that our own galactic center may host an IMBH in the inner pc \citep[e.g.,][]{Hansen+03,Maillard+04,Grkan+20,Gualandris+09,Chen+13,Generozov+20,Fragione+20,Zheng+20,Naoz+20,Gravity+20}.



Several IMBH formation channels have been suggested in the literature. For example, IMBHs may have a cosmological origin, forming in the early universe either as a result of the very first stars \citep[e.g.,][]{MadauRees01,Schneider+02,Johnson+07,Valiante+16} or from direct collapse of accumulated gas \citep[e.g.,][]{Begelman+06,Yue+14,Ferrara+14,Choi+15,Shlosman+16}. These high redshift IMBHs would need to survive galaxy evolution and mergers to present day \citep[e.g.,][]{Rashkov+14}, with significant effects on their stellar and even dark matter surroundings \citep[e.g.,][]{Bertone+09,Chen+13,Bringmann+12,Eda+13,Naoz+14,Naoz+19}. Another popular formation channel relies on the coalescence of many stellar-mass black holes, which may seed objects as massive as SMBHs \citep[e.g.,][]{Kroupa+20}. IMBHs may form in the centers of globular clusters, where few-body interactions lead to the merger of stellar-mass BHs \citep[e.g.,][]{OLeary+06,Gurkan+06,Blecha+06,Freitag+06,Umbreit+12,Rodriguez+18,Rodriguez+19,FragioneLoebRasio20}. Other formation mechanisms invoke successive collisions and mergers of massive stars  \citep[e.g.,][]{Ebisuzaki+01,PortegiesZwart&McMillan02,PortegiesZwart+04,Freitag+06,Sakurai+17,Kremer+20,Gonzalez+21,DiCarlo+21,Das+21b,Das+21,Escala21}.

The main obstacle to sequential BH mergers in clusters is that the merger recoil velocity kick often exceeds the escape velocity from the cluster \citep[e.g.,][Rom \& Sari, in prep.]{Schnittman+07,Centrella+10,OLeary+06,Baibhav+20}.
However, nuclear star clusters at the centers of galaxies do not encounter this problem.
For example, \citet{Fragione+21} explore repeated BH-BH mergers in nuclear star clusters without a SMBH. They considered BH binary-single interactions, binary BH GW merger, and GW merger recoil kicks. The post-kick merger product sinks back towards the cluster center over a dynamical friction timescale. Using this approach, they showed that $10^3-10^4$~M$_\odot$ IMBHs can form efficiently over the lifetime of a cluster.

However, as discussed in Section~\ref{sec:Collisions}, direct BH-star collisions are much more frequent than BH-BH collision in galactic nuclei, making the former a promising channel for BH growth. In an N-body study of young star clusters, \citet{Rizzuto+22} find that BH-star collisions are a main contributor to the formation of BHs in the mass gap and IMBHs. In a similar vein, \citet{Stone+17} demonstrate that massive BHs can form from repeated tidal encounters between stars and BHs. More generally, several studies have explored the role of collisions in a GN, with implications for the stellar and red giant populations \citep[e.g.,][]{DaleDavies,Dale+09,Balberg+13,Mastrobuono-Battisti+21}. We propose that IMBHs can form naturally within the central pc of a galactic center through repeated collisions between BHs and {\it main sequence stars}. During a collision, the BH can accrete some portion of the star's mass. Over many collisions, it can grow appreciably in size.
We demonstrate that this channel can create IMBHs with masses as large as $10^{4}$~M$_\odot$, 
an upper limit that depends on the density profile of the surrounding stars and the efficiency of the accretion.

The paper is structured as follows: we describe relevant physical processes and our approach in Section~\ref{sec:method}. In particular, we provide an overview of collisions in Section~\ref{sec:Collisions} and present our statistical approach in Section~\ref{sec:statistical_coll}. Section~\ref{sec:Accretion} discusses our treatment of the mass growth with each collision and presents analytic solutions to our equations in two different regimes, efficient collisions 
and inefficient collisions 
We compare these solutions to our statistical results. Sections~\ref{sec:GWinspiral} and \ref{sec:EMRIs} discuss implications for GW merger events between IMBHs and the SMBH. We then incorporate relaxation processes and discuss the subsequent results in Section~\ref{sec:Relaxation}. Finally, we discuss and summarize our findings in Section~\ref{sec:predictions}.

\section{Methodology} \label{sec:method}
We consider a population of stellar mass BHs embedded in a cluster of $1$~M$_\odot$ stars. When stars and BHs collide, the BHs can accrete mass. The growth rate depends on the physical processes outlined below. We use a statistical approach to estimate the stellar encounters and final IMBH masses.

\subsection{Physical Picture} \label{sec:ICs}

We consider a population of BHs within the inner few parsecs of the SMBH in a galactic nucleus (GN).
We assume that the BH mass distribution follows that of the stars from which they originate, a Kroupa initial mass function $dN/dm \propto m^{-2.35}$. While this choice represents a gross oversimplification, it has very little bearing on our final results. Future work may address the particulars of the BH mass distribution, but we do not expect that it will significantly alter the outcome. The upper and lower limits of the BH mass distribution are $5$ and $50 \, M_\odot$, respectively. We select the upper limit to encompass the range of upper bounds predicted by stellar evolution models, which vary between $40$ and $125 \, M_\odot$ depending on the metallicity \citep{Heger+03,Woosley+17,SperaMapelli17,LimongiChieffi,Belczynski20,Renzo+20}. We assume that the orbits of the BHs follow a thermal eccentricity distribution. We draw their semimajor axes, $a_\bullet$, from a uniform distribution in log distance, $dN/d(\log r)$ being constant. While this distribution is not necessarily representative of actual conditions in the GN, we use it to build a comprehensive physical picture of BH growth at all distances from the SMBH, including within $0.01$~pc. Otherwise, the innermost region of the GN would be poorly represented in our sample. We consider other observationally motivated distributions in Section~\ref{sec:Relaxation}, but reserve a more detailed examination of the distribution's impact for future work.

\subsection{Direct Collisions} \label{sec:Collisions}

BHs in the GN can undergo direct collisions with other objects. The timescale for this process, $t_{\rm coll}$, can be estimated using a simple rate calculation: $t_{\rm coll}^{-1} = n \sigma A$, where $n$ is the number density of objects, $\sigma$ is the velocity dispersion, and $A$ is the cross-section. We use the collision timescale from \citet{Rose+20}:
\begin{eqnarray} \label{eq:t_coll_main_ecc}
     t_{\rm coll}^{-1} &=& \pi n(a_\bullet) \sigma(a_\bullet) \nonumber \\ &\times& \left(f_1(e_\bullet)r_c^2 + f_2(e_\bullet)r_c \frac{2G(m_{BH}+m_\star)}{\sigma(a_\bullet)^2}\right)\ .
\end{eqnarray}
where $G$ is the gravitational constant and $r_c$ is the sum of the radii of the interacting objects, a black hole with mass $m_{BH}$ and a star with mass $m_\star$.
Detailed in \citet{Rose+20}, $f_1(e_\bullet)$ and $f_2(e_\bullet)$ account for the effect of the eccentricity of the BH's orbit about the SMBH on the collision rate, while $n$ and $\sigma$ are simply evaluated at the semimajor axis of the orbit (see below). Note that this timescale equation includes the effects of gravitational focusing, which enhances the cross-section of interaction.

\begin{figure}
	\includegraphics[width=\columnwidth]{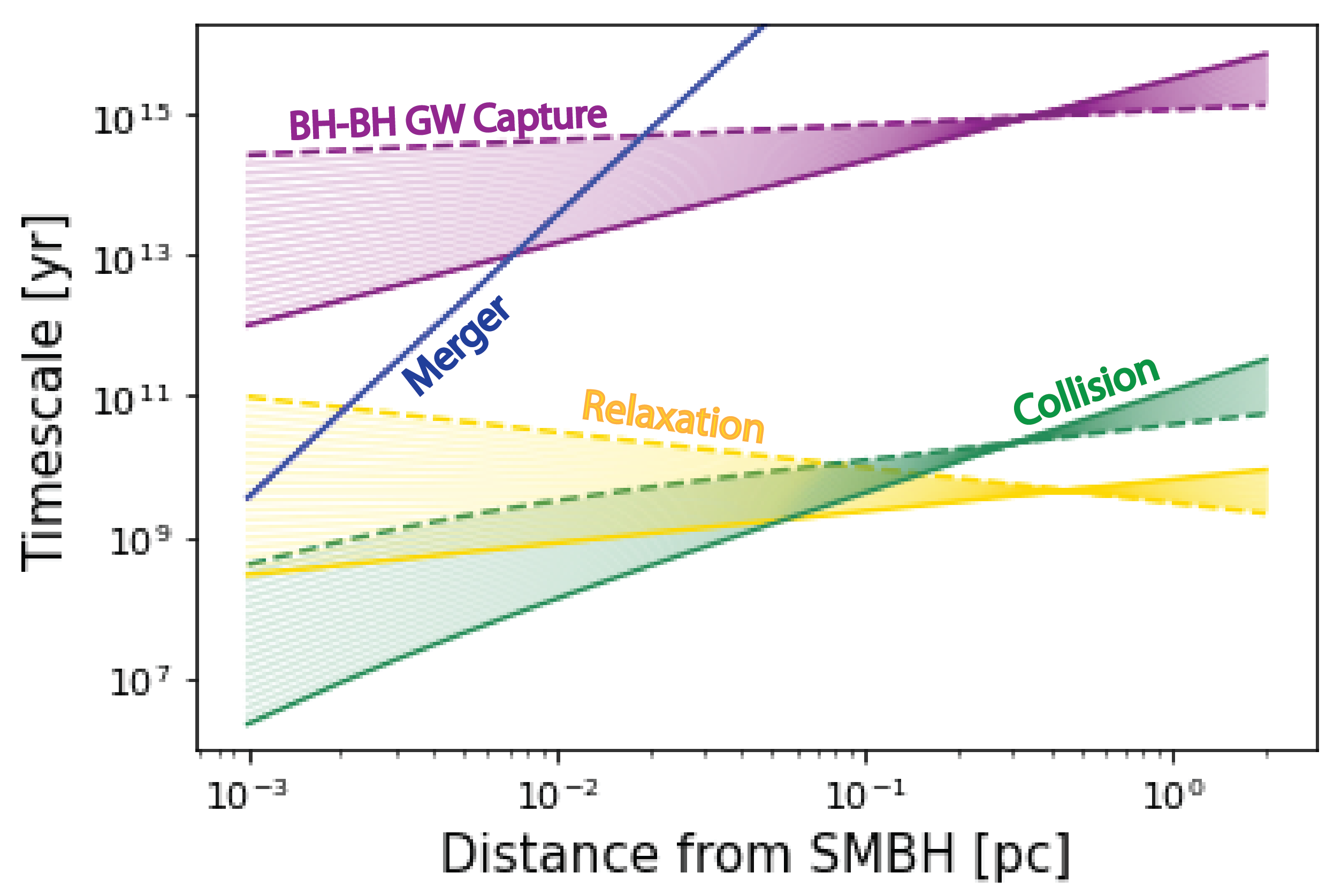}
    \caption{We plot the relevant timescales, including collision (green), relaxation (gold), and BH-BH GW capture (purple), for a single BH in the GN as a function of distance from the SMBH. For the collision timescale, we assume the BH is on a circular orbit. The timescales depend on the density, so we adopt a range of density profiles, bounded by $\alpha = 1$ (dashed curve) to $\alpha = 2$ (dark, solid curve). The dark blue line represents the time for a $10^5$~M$_\odot$ BH to merge with the SMBH through GW emission.}
    \label{fig:timescales}
\end{figure}

Assuming a circular orbit for simplicity, we plot the timescale for a BH orbiting in the GN to collide with a $1 \, M_\odot$ star as a function of distance from the SMBH in Figure~\ref{fig:timescales}.\footnote{We note that the eccentricity has a very minor effect on the collision timescale \citep{Rose+20}.} As this timescale depends on the density of surrounding stars, we adopt a density profile of the form:
\begin{eqnarray} \label{eq:density}
    \rho(r_\bullet) = \rho_0 \left( \frac{r_\bullet}{r_0}\right)^{-\alpha} \ , 
\end{eqnarray}
where $r_\bullet$ denotes the distance from the SMBH. We adopt a SMBH mass of $4\times 10^6$~M$_\odot$ such that our fiducial GN matches our own galactic center \citep[e.g.,][]{Ghez+05,Genzel+03}. In this case, the normalization in Eq.~(\ref{eq:density}) is $\rho_0 = 1.35 \times 10^6 \, M_\odot/{\rm pc}^3$ at $r_0 = 0.25 \, {\rm pc}$ \citep{Genzel+10}. Additionally, in Eq.~(\ref{eq:density}), $\alpha$ gives the slope of the power law. We assume that a uniform population of solar mass stars account for most of the mass in the GN, making the stellar number density:
\begin{eqnarray}
    n(r_\bullet) = \frac {\rho(r_\bullet)}{1 \, M_\odot} \ . \label{eq:density_stars}
\end{eqnarray}
The collision timescale also depends on the velocity dispersion, which we express as:
\begin{eqnarray}\label{eq:sigma}
    \sigma(r_\bullet) = \sqrt{ \frac{GM_{\bullet}}{r_\bullet(1+\alpha)}},
\end{eqnarray}
where $\alpha$ is the slope of the density profile and $M_{\bullet}$ denotes the mass of the SMBH \citep{Alexander99,AlexanderPfuhl14}. As mentioned above,
Eq.~(\ref{eq:t_coll_main_ecc}) depends on the sum of the radii of the colliding objects, $r_c$. We take $r_c = 1 \, R_{\odot}$ because these interactions involve a BH and a star, and the former has a much smaller physical cross-section. For example,  the Schwarzschild radius of a $10 \, M_\odot$ BH is only $30$ km, or $4.31 \times 10^{-5} \, R_\odot$. For this reason, direct collisions between compact objects are very rare and not included in our model.


We note that direct collisions between BHs, via GW emission, were shown to be efficient in nuclear star clusters without SMBHs \citep[e.g.,][]{PortegiesZwart,OLeary+06,RodriquezChatterjee16}. However, in the GN, star-BH collisions are much more frequent than direct BH-BH collisions. As depicted in Figure \ref{fig:timescales}, the star-BH collision timescale for a range of density profiles is many orders of magnitude shorter than the BH-BH GW collision timescale \citep[for the relevant equations, see][for example]{OLeary+09,Gondan+18a}. Thus, we expect that star-BH collisions will be the main driver of IMBH growth in the GN.

\subsection{Statistical Approach to Collisions} \label{sec:statistical_coll}

We simulate the mass growth of a population of BHs with initial conditions detailed in Section~\ref{sec:ICs}. Over an increment $\Delta t$ of $10^6$~yr, we calculate the probability of a collision occurring, given by $\Delta t/t_\mathrm{coll}$. This choice of $\Delta t$ is motivated by our galactic center's star formation timescale \citep[e.g.,][]{Lu+09}, allowing for regular replenishment of the stellar population in the GN. We have checked that the results are not sensitive to this choice of $\Delta t$, omitted here to avoid clutter. We draw a number between $0$ and $1$ using a random number generator. If that number is less than or equal to the probability, we increase the BH's mass by $\Delta m$, the mass that the BH is expected to accrete in a single collision (see Section~\ref{sec:Accretion} for details). We recalculate the collision timescale using the updated BH mass and repeat this process until the time elapsed equals the simulation time of $10$~Gyr\footnote{ Closer to the SMBH, $\Delta t$ may exceed the collision timescale by a factor of a few for steep density profiles. We include a safeguard in our code which takes the ratio $t_\mathrm{coll}/\Delta t$ and rounds it to the nearest integer. We take this integer to be the number of collisions and increase the BH mass accordingly.}.


\subsection{Mass Growth} \label{sec:Accretion}

When a BH collides with a star, it may accrete material and grow in mass. The details of the accretion depend on the relative velocity between the BH and star. For simplicity, this calculation assumes that the two objects experience a head on collision, with the BH passing through the star's center. We begin by considering the escape velocity from the BH at the star's outermost point, its surface, which corresponds to the maximum impact parameter $1$~R$_\odot$. Qualitatively, one might expect that the BH could capture the entire star (i.e., $\Delta m\sim 1$~M$_\odot$) if the relative velocity is smaller than the escape velocity from the BH at this point. However, in the vicinity of the SMBH, the dispersion velocity of the stars may be much larger than the escape velocity from the BH at the star's surface. In this case, the BH captures a ``tunnel'' of material through the star. This tunnel has radius equal to the Bondi radius and length approximately $1 \, R_\odot$.  For the purposes of this study, we assume that the BH accretes all of the material that it captures. The details of the accretion are uncertain, however, and it may be much less efficient than our results imply. We discuss accretion in Section~\ref{sec:Winds}.

To estimate $\Delta m$,  we begin with the Bondi-Hoyle accretion rate, $\dot{m}$, given by:
\begin{eqnarray} \label{eq:b_GWcap}
\dot{m} = \frac{4 \pi G^2 m_\mathrm{BH}^2 \rho_\mathrm{star}}{\left(c_s^2+\sigma^2\right)^{3/2}} \ ,
\end{eqnarray}
where $c_s$ is the speed of sound in the star and $\rho_\mathrm{star}$ is its density \citep[e.g.,][see latter for a review]{Bondi52,BondiHoyle,Shima85,BondiHoyleOverview}.  We approximate the density as $1 \, M_\odot/(4 \pi R_\odot^3/3)$ and take the conservative value of $c_s = 500~km~s^{-1}$, which is consistent with the sound speed inside a $1~M_\odot$ star \citep{Christensen-Dalsgaard+96} and allows us to set a lower limit on $\Delta m$.
To find $\Delta m$, at each collision, we have: 
\begin{equation} \label{eq:deltam}
    \Delta m = {\rm min}(\dot{m}\times t_{\rm \star, cross},1~{\rm M}_\odot) \ ,
\end{equation}
where $t_{\rm \star, cross} \sim R_\odot/\sigma$ is the crossing time of the BH in the star. We take the minimum between  $\dot{m}\times t_{\rm \star, cross}$ and $1~{\rm M}_\odot$ because the BH cannot accrete more mass than one star at each collision.

\begin{figure}
	\includegraphics[width=\columnwidth]{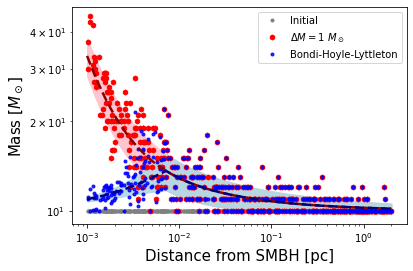}
    \caption{We consider an example that highlights the mass growth as a function of distance from the SMBH.  Grey dots represent the initial masses and distances from the SMBH of the BHs involved in the simulation. For simplicity, we set the inital mass equal to $10 \, M_\odot$ for all of the BHs. Assuming the density profile of stars has $\alpha = 1$, we consider two cases: BHs accrete all of the star's mass during a collision (red) and only a portion of the star's mass is accreted during a collision given by Eq.~\ref{eq:deltam} (blue). The latter case results in less growth closer to the SMBH where the velocity dispersion becomes high. The shaded regions and dashed lines represent the analytical predictions detailed in Section~\ref{sec:Accretion}.} \label{fig:BHaccretion}
\end{figure}

Figure~\ref{fig:BHaccretion} juxtaposes the expected growth using Bondi-Hoyle-Lyttleton accretion (blue small points) with a much simpler model in which the BH accretes the star's entire mass, $1 \, M_\odot$ (red large points). Both examples start with identical populations of $10 \, M_\odot$ BHs (grey) and simulate growth through collisions using a statistical approach. As the BHs grow, the collision timescale, which depends on $m_{BH}$, decreases. Simultaneously, $\Delta m$, which also depends on $m_{BH}$, increases. The result is exponential growth (see discussion and details surrounding Eq.~(\ref{eq:final_mass})).
In Figure~\ref{fig:BHaccretion}, however, the simulations assume $\alpha = 1$ for the stellar density profile, ensuring the collision timescale is long compared to the simulation time, $10$~Gyr. Therefore, the BHs grow slowly, and their final masses can be approximated using the following equation:
\begin{eqnarray} \label{eq:final_mass_bondi}
m_\mathrm{final}({t_{\rm coll}\to {\rm const.}}) 
&=&   m_\mathrm{initial} + \Delta m \frac{T}{t_\mathrm{coll}} \ ,
\end{eqnarray}
in which $T$ represents the simulation time and $\Delta m$ and $t_\mathrm{coll}$ remain constant, approximated as their initial values.

This equation is plotted in Figure~\ref{fig:BHaccretion} for both cases, $\Delta m = 1 \, M_\odot$ (red) and $\Delta m$ from Bondi-Hoyle-Lyttleton accretion (blue), and the curves coincide with the corresponding simulated results. The shaded regions represent one standard deviation from Eq.~(\ref{eq:final_mass_bondi}), calculated using the square root of the number of collisions, $T/t_\mathrm{coll}$. As indicated by the results in red, in the absence of Bondi-Hoyle-Lyttleton accretion, the BHs closest to the SMBH experience the most growth because they have shorter collision timescales. However, Bondi-Hoyle-Lyttleton accretion becomes important closer to the SMBH, where the velocity dispersion is large {compared with the stars' escape velocity}, and curtails the mass growth for BHs in this region. Outside of $10^{-2}$~pc, a BH consumes the star's entire mass: the accretion-limited $\Delta m$ governed by Eq.~(\ref{eq:final_mass_bondi}) is greater than or equal to the star's mass.

\begin{figure*}
	\includegraphics[width=0.98\textwidth]{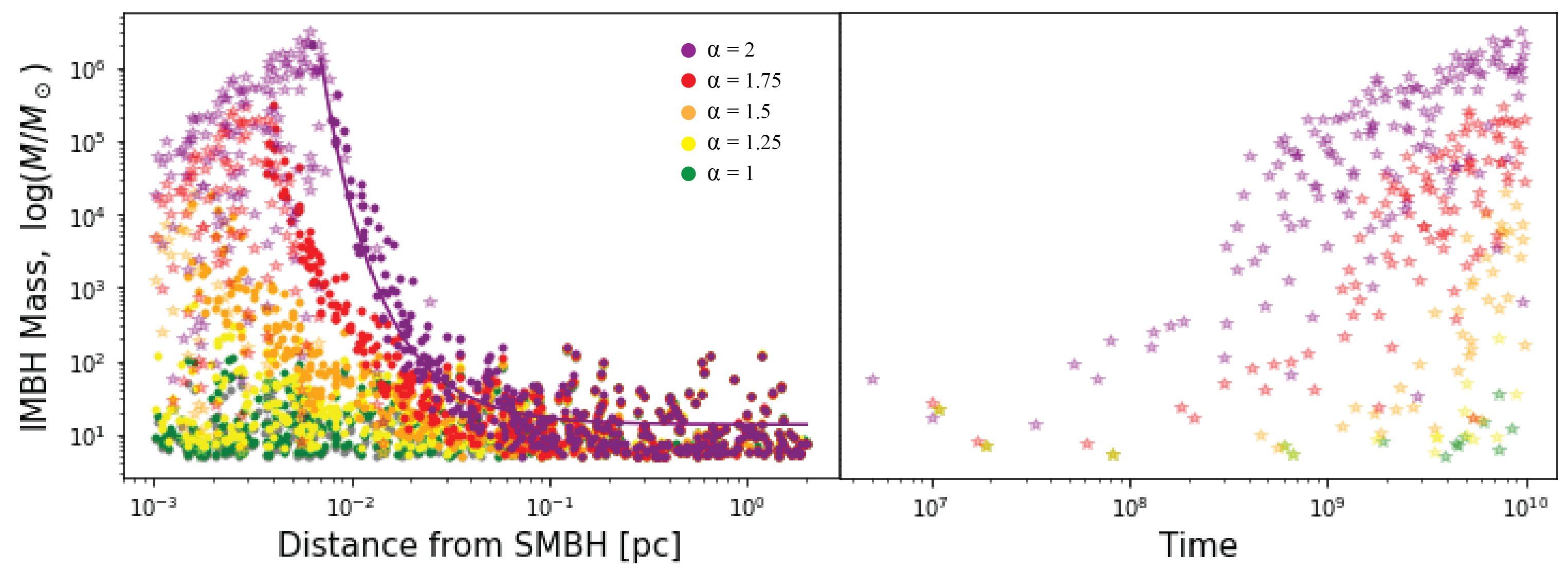}
    \caption{On the right, we plot final masses of 500 BHs using different values of $\alpha$ in the density profile, shallow ($\alpha = 1$) to cuspy ($\alpha = 2$). For the latter case, the purple line shows the analytical result from Eq.~\ref{eq:final_mass}, taking $m_\mathrm{initial}$ to be the average mass of the population. Faded stars indicate BHs that merged with the SMBH through GWs. On the left, we plot the masses and merger times of these BHs.}
    \label{fig:alpha}
\end{figure*}

Eq.~\ref{eq:final_mass_bondi} does not apply for other values of $\alpha$. When the collision timescale is shorter, corresponding to a larger index $\alpha$ in the density profile (see Figure \ref{fig:timescales}), the growth is very efficient and $\Delta m$ quickly approaches $1 \, \mathrm{M}_\odot$. Consequently, while we can now assume $\Delta m = 1 \, \mathrm{M}_\odot$, we can no longer assume the collision timescale is constant. The final mass grows exponentially as a result. For $\Delta m = 1 \mathrm{M}_\odot$, the general solution is reached by solving
the differential equation $dm/dt = 1 \, M_\odot/t_\mathrm{coll}(m)$, which gives: 
\begin{eqnarray} \label{eq:final_mass}
     m_{\mathrm{final}}(\Delta m\to 1 \, \mathrm{M}_\odot) &=& -A+\left(m_\mathrm{initial}+A\right)e^{CT}
\end{eqnarray}
where $A = \sigma^2 R_\mathrm{star}/G$ and $C = 2\pi G n_\mathrm{star} R_\mathrm{star}/\sigma$.
As an example, we plot this curve in purple for the $\alpha = 2$ case, in Figure \ref{fig:alpha}, which agrees with the simulated masses.

\subsection{Uncertainties in Accretion} \label{sec:Winds}

We note that the $\Delta M$ calculated in this proof-of-concept study assumes that the BH accretes all of the material that it captures. Estimating the true fraction of the material accreted by the BH is very challenging; this complex problem requires numerically solving the generalized GR fluid equations with cooling, heating, and radiative transfer, etc. and remains an active field of research \citep[e.g.,][]{BlandfordBegelman99,ParkOstriker+01,Narayan+03,Igumenshchev+03,Ohsuga+05,Yuan+12,Jiang+14,McKinney+14,Narayan+22}. Heuristically, if a collision between a BH and a star results in an accretion disk, the disk’s viscous timescale may be as low as days. The resultant luminosity can unbind most of the captured material, though details such as the amount accreted and peak luminosity remain uncertain (e.g., \citet{Yuan+12,Jiang+14}, see also the discussion in \citet{Stone+17}, \citet{Rizzuto+22}, and \citet{Kremer+22}). The question becomes whether or not a BH can still accumulate significant amounts of mass over many collisions even if it accretes very little in a single one. We explore the viability of our channel using a physically motivated inefficient accretion model. Several studies have invoked momentum-driven winds in BH accretion \citep[e.g.,][]{Murray+05,Ostriker+10,Brennan+18}. We thus estimate the fraction of captured mass accreted to be approximately $v_{esc}/(c \eta)$, where $v_{esc}$ is the escape velocity from the BH at $1$~R$_\odot$ and $\eta$ is the accretion efficiency at the ISCO. We take $\eta$ to be $0.1$ \citep[e.g.,][]{YuTremaine02}. This expression for the fraction accreted is consistent with \citet{Kremer+22} equation 19 for $s = 0.5$, which is a reasonable value for $s$, a free parameter between $0.2$ and $0.8$. We discuss the results of the momentum-driven winds estimate in Section~\ref{sec:predictions}. We note that the accretion process may be more efficient than this estimate implies if, for example, jets or other instabilities result in the beaming of radiation away from the captured material \citep[e.g.,][]{BlandfordZnajek77,Begelman79,DeVilliers+05,McKinneyGammie,McKinney06,Igumenshchev+08,Begelman12,Begelman12b,McKinney+14}.

\subsection{GW Inspiral} \label{sec:GWinspiral}

When a BH is close to the SMBH, GW emission can circularize and shrink its orbit. We implement the effects of GW emission on the BH's semimajor axis and eccentricity following \citet{Peters+63}. The characteristic timescale to merge a BH with an SMBH is given by: 
\begin{eqnarray}
    t_{GW} &\approx&  2.9\times 10^{12} ~{\rm yr}\left(\frac{M_\bullet}{10^6~{\rm M}_\odot} \right)^{-1} \left(\frac{m_{BH}}{10^6~{\rm M}_\odot} \right)^{-1}\nonumber \\ &\times& \left( \frac{M_\bullet+m_{BH}}{2 \times 10^6~{\rm M}_\odot} \right)^{-1} \left(\frac{a_\bullet}{10^{-2}~{\rm pc}} \right)^4 \nonumber \\ 
    &\times& f(e_\bullet)(1-e_\bullet^2)^{7/2} \ ,
\end{eqnarray}
where $f(e_\bullet)$ is a function of $e_\bullet$. For all values of $e_\bullet$, $f(e_\bullet)$ is between $0.979$ and $1.81$ \citep[][]{Bla+02}. We plot this timescale for a $1 \times 10^5 \, M_\odot$ BH in Figure~\ref{fig:timescales} in blue.

In our simulations, we assume a BH has merged with the SMBH when the condition $t_{GW} < t_\mathrm{elapsed}$ is met. When this condition is satisfied, we terminate mass growth through collisions for that BH.\footnote{For comparison, we also incrementally changed the semimajor axis and eccentricity from GW emission following the equations in \citet{PetersMathews63}. This method leads to a slight increase in the final IMBH masses because it accounts for the collisions that take place while the orbit is gradually shrinking.}

\subsection{IMBH growth}

As detailed above, BH-stellar collisions can increase the BH masses as a function of time. Here, we examine the sensitivity of the BH growth to the density power law. From Eq.~(\ref{eq:t_coll_main_ecc}), it is clear that the growth rate depends on the stellar density profile, governed by the index $\alpha$. We expect that higher values of $\alpha$, or steeper profiles, will result in more efficient mass growth. In Figure~\ref{fig:timescales}, larger values of $\alpha$ lead to collision timescales in the GN's inner region, inwards of $0.25$~pc, that are much smaller that the $10$~Gyr simulation time. Figure \ref{fig:alpha} confirms this expectation. It depicts the mass growth of a uniform distribution of BHs with initial conditions detailed in Section~\ref{sec:ICs} for five $\alpha$ values, spanning $1$ (green) to $2$ (purple). The most massive IMBHs form inwards of $0.25$~pc for the $\alpha = 2$ case.

\subsection{Gravitational Wave Mergers and Intermediate and Extreme Mass Ratio Inspiral Candidates} \label{sec:EMRIs}
Towards the SMBH, efficient collisions can create BHs massive enough to merge with the SMBH through GWs. Following the method detailed in Section~\ref{sec:GWinspiral}, when a given BH meets the criterion $t_\mathrm{GW} < t_\mathrm{elapsed}$, we mark it as merged with the SMBH. We assume that at this point the dynamics of the BH will be determined by GW emission, shrinking and circularizing the BHs orbit until it undergoes an extreme or intermediate mass ratio inspiral (EMRI and IMRI, respectively). The righthand plot in Figure~\ref{fig:alpha} shows the BH masses versus time of merger. It is interesting to note that even in the absence of relaxation processes, which are often invoked to explain the formation of EMRIs, EMRIs and notably IMRIs can form in this region.


\subsection{Two Body Relaxation Processes} \label{sec:Relaxation}

A BH orbiting the SMBH experiences weak gravitational interactions with other objects in the GN. Over a relaxation time, these interactions alter its orbit about the SMBH. The two-body relaxation timescale for a single-mass system is: 
\begin{eqnarray} \label{eq:t_rlx}
t_{\rm relax} = 0.34 \frac{\sigma^3}{G^2 \rho \langle M_\ast \rangle \ln \Lambda_{\rm rlx}},
\end{eqnarray}
where $\ln \Lambda_{\rm rlx}$ is the Coulomb logarithm and $\langle M_\ast \rangle$  is the average mass of the surrounding objects, here assumed to be $1 \, M_\odot$ \citep[][Eq.~(7.106)]{Spitzer1987,BinneyTremaine}. This equation represents the approximate timescale for a BH on a semi-circular orbit to change its orbital energy and angular momentum by order of themselves. The BH experiences diffusion in its angular momentum  and energy as a function of time  \citep[depending on the eccentricity of the orbit, this process can be more efficient][]{FragioneSari18,SariFragione19}. Relaxation can cause the orbit of an object in a GN to reach high eccentricities. If the object is a BH, it can spiral into the SMBH and form an EMRI, while a star can be tidally disrupted by the SMBH \citep[e.g.][]{MagorrianTremaine99,WangMerritt04,HopmanAlexander05,AharonPerets16,StoneMetzger16,Amaro-Seoane18,SariFragione19,Naoz+22}. The relaxation process is therefore crucial to our study.
In Figure~\ref{fig:timescales}, we plot the relaxation timescale in gold for a range of $\alpha$. We note that the \citet{BahcallWolf76} profile, $\alpha=7/4$, corresponds to zero net flux and therefore does not preferentially migrate objects inward.

Additionally, because BHs are more massive on average than the surrounding objects, they are expected to segregate inwards in the GN \citep[e.g.,][]{Shapiro+78,Cohn+78,Morris93,MiraldaEGould00,Baumgardt+04}. They sink toward the SMBH on the mass segregation timescale, $t_{\rm seg}\approx \langle M_\ast \rangle/m_{\rm BH} \times t_{\rm relax}$ \citep[e.g.,][]{Spitzer1987,Fregeau+02,Merritt06}, which is typically an order of magnitude smaller than the relaxation timescale plotted in Figure~\ref{fig:timescales}. 

We incorporate relaxation processes 
by introducing a small change in the BH's energy and angular momentum each time it orbits the SMBH. We apply a small instantaneous velocity kick to the BH, denoted as $\Delta v$. We draw $\Delta v$ from a Guassian distribution with average of zero and a standard deviation of $\Delta v_{rlx}/\sqrt{3}$, where $ \Delta v_{rlx} = v_\bullet\sqrt{ {P_\bullet}/{t_{rlx}}}$\ \citep[see][for an approach to changes in the angular momentum]{Bradnick+17}.  The new orbital parameters can be calculated following \citet{Lu+19}, and see \citet{Naoz+22} for the full set of equations.


We account for the effects of relaxation processes, including mass-segregation, using a multi-faceted approach. We begin by migrating each BH towards the center over its mass-segregation timescale, shifting it incrementally inward such that its orbital energy changes by order of itself within the segregation timescale.

As the BHs segregate down the potential well, their abundance with respect to stars increases, until at some turnover radius, BHs become the dominant source of scattering for both black holes and stars. Within this radius, BH self-interaction dominates over two-body scatterings with the now rarer main-sequence stars. The BHs will then settle onto a Bahcall-Wolf profile, while the stars may follow a shallower profile, with approximately $n_\star \propto r^{-1.5}$, inwards of the transition radius (Linial \& Sari in prep.).

Therefore, after the initial mass segregation, we allow the BHs to begin diffusing over a relaxation timescale, their orbital parameters changing slowly through a random process. In this random process, some of the BHs may migrate closer to the SMBH. We terminate mass growth when the BH enters the inner $200$~au of the GN, within which the density of stars is uncertain. This cutoff is based on the $120$~au pericenter of S0-2, the closest known star to the SMBH \citep[e.g.,][]{Ghez+05}.

Another physical process that causes inward migration is dynamical friction. 
A cursory derivation based on the dynamical friction equations described in \citet{BinneyTremaine} reveals the process to have a similar timescale to mass segregation. If a BH diffuses to a distance greater than $2$~pc from the SMBH, exiting the sphere of influence, we have it sink inwards, back towards the center, over a dynamical friction timescale. After one dynamical friction timescale has passed, we restart diffusion.

We note that our prescription ignores self-interactions between the BHs. As mentioned above, as the BHs sink towards the SMBH, their concentration in the inner region of the GN increases, allowing them to dominate the scattering. We reserve the inclusion of these interactions for future study.



\begin{figure*}
	\includegraphics[width=0.98\textwidth]{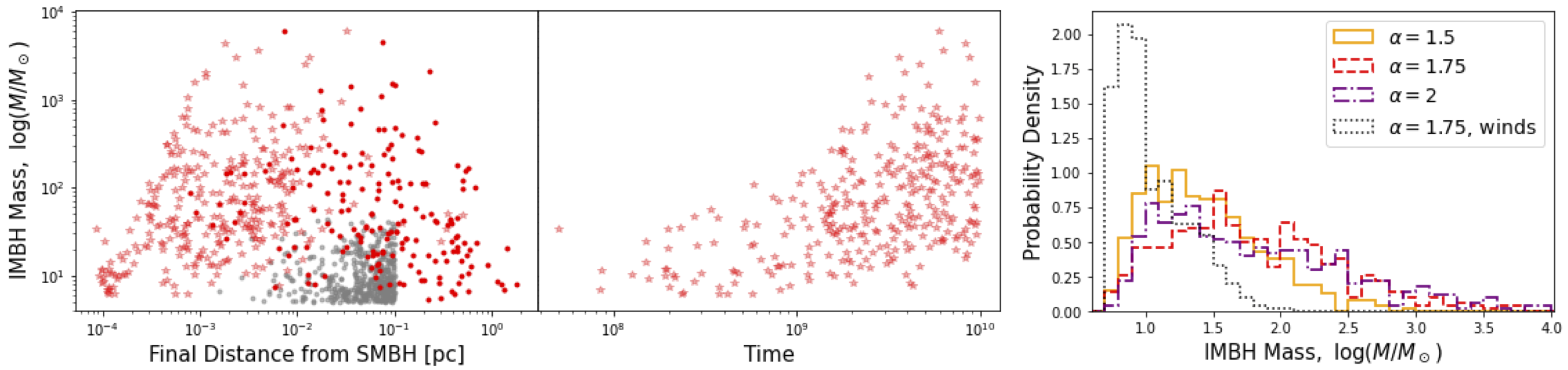}
    \caption{Similar to Figure~\ref{fig:alpha}, we plot the initial masses versus initial distance (grey) and final mass versus final distance (red) for 500 BHs. This simulation includes relaxation processes, including mass segregation, diffusion, and dynamical friction. We assume $\alpha = 1.75$ for the GN density profile. Faded stars represent BHs that merge with the SMBH. As a result of inward migration, BHs merge more quickly with the SMBH, before they can become as massive as those in Figure~\ref{fig:alpha}. Additionally, more BHs become EMRIs and IMRIs.
    Additionally, in the third panel, we show a histogram of the simulated IMBH masses for two different values of $\alpha$, $1.5$ (orange, solid), $\alpha$, $1.75$ (red, dashed), and $2$ (purple, dash-dotted), accounting for relaxation processes. We also show the results for a simulation with $\alpha = 1.75$ that accounts for momentum-driven winds (black, dotted). Despite the substantially reduced accretion, BHs in the mass gap still form.}
    \label{fig:rlx_BW}
\end{figure*}

\subsection{Effect of Relaxation Processes}
As depicted in Figure \ref{fig:rlx_BW}, two-body relaxation processes result in more EMRIs and IMRIs events. These processes allow BHs that begin further from the SMBH to migrate inwards and grow more efficiently in mass. However, it also impedes the growth of BHs that are initially closer to the SMBH by allowing them to diffuse out of the inner region where collisions are efficient. As can be seen in Figure \ref{fig:rlx_BW}, the net result is that more BHs grow, but the maximum mass is lower compared to the scenario that ignores two-body relaxation. The histogram in Figure~\ref{fig:rlx_BW} presents the final BH mass distributions for different power law indices $\alpha$. As expected, the two-body relaxation suppresses the $\alpha$ dependence highlighted in Figure \ref{fig:alpha}. In fact, using a KS test, we find that we cannot reject the hypothesis that the two distributions were drawn from the same sample for the $\alpha = 1.75$ and $\alpha=2$ results. Interestingly, a BH mass IMF with an average of $10$~M$_\odot$ leads to a final distribution with an average of $\sim 200$~M$_\odot$ and a median of $\sim 45$~M$_\odot$, which lies within the mass gap.

\section{Discussion and Predictions} \label{sec:predictions}


We explore the feasibility of forming IMBHs in a GN through successive collisions between a stellar-mass BH and main-sequence stars. Taking both a statistical and analytic approach, we show that this channel can produce IMBHs efficiently with masses as high as $10^{3-4}$~M$_\odot$ and may result in many IMBH-SMBH mergers (intermediate-mass ratio inspirals, or IMRIs) and EMRIs.

As the stellar mass BH collides with a star, the BH will grow in mass. The increase may equal star's entire mass if the relative velocity is smaller than the escape velocity from the BH at $1$~R$_\odot$. However, near the SMBH, the velocity dispersion may be larger than the escape velocity from the BH at the star’s radius. In this limit, the BH captures a “tunnel” of material through the star, estimated using Bondi-Hoyle-Lyttleton accretion. In our statistical analysis, we account for Bondi-Hoyle-Lyttleton accretion and find that BHs outside of $10^{-2}$~pc from the SMBH can capture the entire star (see Figure \ref{fig:BHaccretion}).

The efficiency of collisions, and therefore IMBH, EMRI, and IMRI formation as well, are sensitive to the underlying stellar density. As shown in Figure~\ref{fig:alpha}, a steeper density profile results in larger IMBHs. This behavior can be understood from the collision timescale's dependence on the stellar density profile. A steeper profile yields shorter collision timescales near the SMBH. However, the inclusion of relaxation processes in the simulations dampens the influence of the stellar density profile by allowing BHs to diffuse into regions of more or less efficient growth. As a result, more BHs grow in mass, but their maximum mass is smaller ($\sim 10^4$~M$_\odot$). Additionally, the final masses have no apparent dependence on distance from the SMBH (see Figure~\ref{fig:rlx_BW}).

Most simulations in our study assume that the BHs accrete all of the mass that they capture. The final BH masses can be taken as an upper limit. We note that the accretion is a highly uncertain process and represents an active field of study \citep[e.g.,][]{BlandfordBegelman99,ParkOstriker+01,Narayan+03,Igumenshchev+03,Ohsuga+05,Yuan+12,Jiang+14,McKinney+14,Narayan+22}. To assess the limits of our model, we also consider a physically motivated accretion model, momentum-driven winds (Section~\ref{sec:Winds}). We present the final mass distribution for momentum-driven winds in Figure~\ref{fig:rlx_BW}. Importantly, we find that BHs within the mass gap still form naturally despite the substantially reduced accretion. About $5\%$ of the BHs grow by $10$ to $100$~M$_\odot$. Furthermore, if we increase this $\Delta M$ estimate by a factor of $2$ (i.e., use $\eta = 0.05$), the simulation produces a $3.5 \times 10^3$~M$_\odot$ IMBH for the same initial conditions. Our proof-of-concept demonstrates that collisions between BH and stars are an important process that should be taken into account in dense places such as a GN.

Mass growth through BH-main-sequence star collisions may act in concert with other IMBH formation channels, such as compact object binary mergers \citep[e.g.,][]{Hoang+18,Stephan+19,Fragione+21,Wang+21}. While in some cases collisions can unbind a binary \citep[e.g.,][]{Sigurdsson&Phinney93,Fregeau+04}, BH binaries can be tightly bound enough to withstand the collisions. Wide binaries may also become unbound due to interactions with the neighboring stars and compact objects \citep[e.g.,][see latter study for the timescale for an arbitrary eccentricity]{BinneyTremaine87,Rose+20}. However, as highlighted in previous studies, a substantial fraction of these binaries may merge due to the Eccentric Kozai Lidov mechanism, leaving behind a single star or a single compact object \citep[e.g.,][]{Stephan+16,Stephan+19,Hoang+18}. Additionally, to be susceptible to evaporation, BH binaries must have a wider configuration. Otherwise, they will be more tightly bound than the average kinetic energy of the surrounding objects and will only harden through weak gravitational interactions with neighboring stars \citep[see for example Figure 6 in][]{Rose+20}.

We note that we assume a steady-state and treat the stars as a reservoir in this model. Future work will take a more nuanced approach to the background stars, whose density as a function of time can be influenced by several factors. Firstly, the relaxation of the stellar population occurs on Gyr timescales. Some studies have suggested that in situ star formation can occur in the Galactic Center as close as 0.04 pc from the SMBH \citep[e.g.,][]{Levin+03,Paumard+06}, and star formation episodes can occur as often as every $\sim 5$~Myr \citep[e.g.][]{Lu+09}. Therefore, we expect that after the first Gyr, stars within $\lesssim 0.01$~pc will be replenished at intervals consistent with the star formation episodes; the infalling populations of stars are separated by $\sim 5-10$~Myr, which is shorter than the collision timescale.

However, star-star collisions may complicate this picture within $\sim 0.01$~pc. As discussed above, regular star formation ensures the BHs always have a stellar population to interact with outside of $\sim 0.01$~pc.\footnote{In fact, the star-star collision timescale is greater than 10 Myr for the entire parameter space, save at $0.001$ pc for larger values of $\alpha$; the BH-star collision timescale plotted in Fig. 1 is the same order of magnitude as the star-star collision timescale.} At $0.01$ pc, however, the kinetic energy during a collision between two $1$~M$_\odot$ stars is larger than their binding energies. Collisions can therefore thin out the stellar populations during the time it takes them to diffuse to these small radii, $\lesssim 0.01$~pc, and may reduce the BH growth in the innermost region. We reserve the inclusion of star-star collisions for future work. We also note that the disruption of binary stars by the SMBH may help replenish the stellar population even as collisions work to deplete it \citep[e.g.,][]{Balberg+13}; when a binary is disrupted, one of the stars is captured on a tightly bound orbit about the SMBH. 

An IMBH may also affect the stellar density profile. As it spirals into the SMBH, it can perturb stellar orbits, and these interactions can lead to hypervelocity stars \citep[e.g.,][]{Baumgardt+06b,LockmannBaumgardt08}. \citet{LockmannBaumgardt08} show that an IMBH can modify an initially steep stellar density profile to become consistent with the flatter cusp observed in the Galactic Center. The stars may then be replenished on $100$~Myr timescales \citep[][]{Baumgardt+06b}. Therefore, after the formation of the first few IMBHs, subsequent BH growth may occur in bursts, coinciding with replenishment of the stars.

While there are many competing dynamical processes that shape the stellar density profile, we stress that $\alpha$ can simply be chosen to encapsulate all of the relevant physics. A value for $\alpha$ that is constrained by observations must already reflect ongoing processes like star-star collisions and replenishment. \citet{Schodel+18} find the observed stellar mass enclosed within 0.01 pc of the Milky Way’s Galactic Center to be approximately $180$~M$_\odot$. This estimate is consistent to order of magnitude with our $\alpha = 1.25$ case. In a simulation like those depicted in Figure 4, which include relaxation, $\alpha = 1.25$ leads to a maximum IMBH mass of $140$~M$_\odot$. Furthermore, while the stellar mass within 0.01 pc may be a few hundred M$_\odot$, \citet{Do+19} and \citet{Gravity+20} set an upper limit on the mass enclosed within the orbit of S0-2 to be about a few thousand M$_\odot$, or $0.1\%$ of the central mass. This upper limit can include mass that was previously in stars but is now in BHs. In that case, the $180$~M$_\odot$ is what remains of the stars, while BHs and IMBHs make up the $\sim 1000$~M$_\odot$ in the innermost region.

Also not included in this study, collisions between the BH and other compact objects will increase the BH growth rate. BH-BH mergers \citep[e.g.,][]{OLeary+09,Fragione+21} and even neutron star BH mergers \citep[e.g.,][]{Hoang+20} become more likely as the BHs increase in mass through stellar collisions. As a result, the BH-BH collision timescale, discussed in Section~\ref{sec:Collisions}, will become relevant to our simulations, allowing the BHs to grow through this channel in addition to stellar collisions. Additionally, this compact object mergers result in GW recoil, which may have a large impact on the dynamics \citep[e.g.,][]{Baibhav+20,Fragione+21}.

The BH's mass growth increases GW emission, which dissipates energy from the orbit. Along with relaxation, GW emission causes BHs to sink towards the SMBH and eventually undergo a merger. As a result, the GN environment is conducive to the formation of EMRIs and IMRIs. The GW emission from EMRIs and IMRIs is expected to be at mHz frequencies, making them promising candidates for LISA to observe. While the exact rate calculation is beyond the scope of this study, the mechanism outlined here seems very promising. 

Our results also suggest that BHs within the mass gap as well as IMBHs likely exist in many galactic nuclei, as well as within our own galactic center. This implication seems to be consistent with recent observational and theoretical studies \citep[e.g.,][]{Hansen+03,Maillard+04,Grkan+20,Gualandris+09,Chen+13,Generozov+20,Fragione+20,Zheng+20,Naoz+20,Gravity+20}.

Lastly, the collisions between stellar mass BHs and stars may contribute to the x-ray emission from our galactic centre \citep[e.g.,][see \citet{Kremer+22} for a discussion of electromagnetic signatures from BH-star collisions]{Muno+05,Muno+09,Hailey+18,Zhu+18,Cheng+18}\footnote{ The connection between the observed X-ray sources at the Galactic Center and tidal capture has been suggested by \citet{Generozov+18}, but see \citet{,Zhu+18,Stephan+19} for alternative channels.}. These interactions, in particular grazing collisions, may also result in tidal disruption events \citep[e.g.,][]{Baumgardt+06,Perets+16,Stone+17,Samsing+19,Kremer+21}. Thus, the process outlined here may produce electromagnetic signatures in addition to GW mergers.





\begin{acknowledgments}
We thank the anonymous referee for useful comments. We also thank Jessica Lu, Fred Rasio, Kyle Kremer, Ryosuke Hirai, Ilya Mandel, and Erez Michaely for useful discussion.

SR thanks the Charles E. Young Fellowship, the Nina Byers Fellowship, and the Michael A. Jura Memorial Graduate Award for support. SR and SN acknowledge the partial support from NASA ATP  80NSSC20K0505. SN thanks Howard and Astrid Preston for their generous support. IL thanks support from the Adams Fellowship. SN and RS thank the Bhaumik Institute visitor program.
This work was performed in part at the Aspen Center for Physics, which is supported by National Science Foundation grant PHY-1607611.

\end{acknowledgments}








\bibliography{sample631}{}
\bibliographystyle{aasjournal}



\end{document}